# The Complexity of Social Coordination

Konstantinos Mamouras, Sigal Oren, Lior Seeman, Lucja Kot, and Johannes Gehrke
Cornell University
Ithaca, NY 14853, USA
{mamouras,sigal,lseeman,lucja,johannes}@cs.cornell.edu

## ABSTRACT

Coordination is a challenging everyday task; just think of the last time you organized a party or a meeting involving several people. As a growing part of our social and professional life goes online, an opportunity for an improved coordination process arises. Recently, Gupta et al. proposed entangled queries as a declarative abstraction for data-driven coordination, where the difficulty of the coordination task is shifted from the user to the database.

Unfortunately, evaluating entangled queries is very hard, and thus previous work considered only a restricted class of queries that satisfy *safety* (the coordination partners are fixed) and *uniqueness* (all queries need to be satisfied).

In this paper we significantly extend the class of feasible entangled queries beyond uniqueness and safety. First, we show that we can simply drop uniqueness and still efficiently evaluate a set of safe entangled queries. Second, we show that as long as all users coordinate on the same set of attributes, we can give an efficient algorithm for coordination even if the set of queries does not satisfy safety. In an experimental evaluation we show that our algorithms are feasible for a wide spectrum of coordination scenarios.

## 1. INTRODUCTION

Since the dawn of humanity coordination has been an important part of social interaction. Where should we gather for the hunt? Who is joining us for war? Where should we have lunch? When should we take our vacation? As technology is evolving, the methods for coordination advance as well; from meeting in person to talking over the phone and eventually emailing, texting and shared calendars. However, as many new applications demonstrate [7, 9], the search for better coordination methods is still in progress.

Coordination has attracted research interest in various areas of computer science. In the field of artificial intelligence there exists extensive work on multiagent systems [14]. Systems researchers have developed communications solutions ranging from low-level mechanisms such as message passing, shared memory, locks and semaphores to higher level abstractions such as transactional memory [11]. The programming languages community has given us languages with support for concurrency and communication, as well as higher level abstractions such the Π calculus [12]. Database constructs such as Sagas [4] could in principle be used for coordination, and solutions for particular scenarios like Web services have also been proposed [3, 13]. However, as discussed in [5], none of these solutions is fully satisfactory for social coordination; for example, most of them do not cleanly separate the coordination specification and mechanism.

Kot et al. [10] were the first to suggest a holistic solution to the problem by augmenting database queries with coordination information, hence shifting the burden of coordination from the users or application designers to the database system itself. They named these augmented queries *entangled queries*. Unlike traditional database queries, these queries are not evaluated in isolation as they may refer to each other and need information from each other to select values. The main challenge in evaluating a set of entangled queries is finding a subset of the queries which is a *coordinating set* – a set of queries such that each query in the set is satisfied and its coordination requirements are also satisfied by queries in the set. Entangled queries have a wide range of application scenarios ranging from travel planning to college students coordinating which classes to take, professionals scheduling joint meetings or players in an MMO game figuring out a battle plan.

Gupta et al. [5] present an algorithm that can find the coordinating set if the queries exhibit two properties: safety and uniqueness. Informally, a set of queries is *safe* if each user specifies exactly the other users she wants to coordinate with. A set of queries is unique if — in order to satisfy one user's coordination requirements — all of the users' coordination requirements must be satisfied. The caveat in this previous approach is that if the queries are not restricted, finding the coordinating set might not be tractable. As the following example shows these two properties are not robust.

**Example 1** *Assume that all the band members of Coldplay want to book tickets from L.A. to Zurich. If each band member specifies the names of the other band members as his coordination partners and no one outside the band asks to fly with one of the band members, then the this set of queries is safe and unique.*

*Now, imagine that Gwyneth Paltrow also submits a request to travel with her husband (who is in the band); the set of queries is no longer unique.*

## 1.1 Contributions

This paper gives a principled treatment of coordination, both from a theoretical and an application perspective. Our results include a clear characterization of various sources of the hardness of coordination as well as practical algorithms that can be tailored to specific application scenarios. We make the following three contributions.

**Pinpointing Hardness.** We start with a crisp separation between the complexity of evaluating conjunctive queries and evaluating entangled queries. We achieve this by reducing 3SAT to the problem of finding a coordinating set of entangled queries in a setting where the database contains only two values. By using such a simple database in the reduction we guarantee that testing for satisfiability of conjunctive queries on the database can always be done in polynomial time. Thus the source of hardness in this reduction is the actual process of finding a set of coordinating queries whose coordination requirements are all satisfied. Previous results did not provide such a clear illumination of the source of hardness. Our hardness results motivate the two practical algorithms which are our next contributions.

**Abolishing Uniqueness.** Our first practical contribution is lifting the uniqueness requirement for general conjunctive entangled queries. This significant expansion of use cases allows us to handle scenarios such as the one stated above, in which all coordination partners are known, but the coordination structure itself is not unique. For instance, in the example above, where there are two potential coordinating sets (Gwyneth and the band members, the band members), our new algorithm (SCC Coordination Algorithm) is guaranteed to find a flight satisfying one of those potential coordinating sets if such a flight exists.

At a high level, our algorithm constructs a small number of subsets of queries which have to be satisfied together. For each subset of queries it issues a single database query to check whether all queries in the subset can be satisfied simultaneously, as was done in Gupta et al. [5]. By the safety property we have that any coordinating set contains at least one of these subsets. Our algorithm's guarantees are quite modest: to find a coordinating set if one exists. However, it is impossible to guarantee much more as we show that finding a coordinating set of maximum size is NP-hard.

**Handling Safety.** Existing solutions restrict the set of coordinating queries in such a way that each user needs to name the friends she wants to coordinate with. However, in many cases this requirement is not desired; consider, for example, the goal of finding a party to go to which at least one of your friends is also going, reserving a flight to a conference with one of your colleagues, or enrolling in a class which one of your friends is also taking. To guarantee that these sets of queries are safe, users need to select coordination partners through a round of out-of-band "pre-coordination" — precisely the scenario entangled queries were designed to avoid. Fortunately, we observe that many coordination scenarios follow a common pattern where users want to coordinate with their friends on a common "thing," such as a party, a flight or a class. Taking this application knowledge into account, we give an efficient algorithm that is able to find a coordinating set even if the set of queries is not safe. We illustrate the idea behind this algorithm with an example.

**Example 2** *Consider a group of Coldplay fans, each one of whom wants to go to a Coldplay concert with at least one of their friends. Since they live in different parts of the world, they cannot take the same flight. However, each fan wants to go with his or her friends to the same location, at about the same time, to see the Coldplay concert. Thus the attributes these users are coordinating on are the flight's destination and the date. They also have an extra requirement that a Coldplay concert should take place at that location, say a day after the flight arrives.*

*Some of the fans might specify a specific location for the concert they want to attend and some might have extra personal requirements which they do not wish to coordinate on, such as the airline.*

When all fans coordinate on the same attributes (destination and date), finding a coordinating set is easy. To find a subset of the fans whose requirements can all be satisfied, we can loop over the Coldplay tour schedule. For each concert we compute a candidate coordinating set which includes all fans that either specify this concert as the concert they want to attend or do not specify a concert at all. For each candidate set we can check for each fan if there exists a flight which will get her to her destination in time for the concert; if not, delete her query and all queries that required it for coordination. If at the end of the process we are left with a subset for which all the coordination requirements are fulfilled, we found a coordinating set.

The beauty of this simple algorithm is that given some specific knowledge of the problem's coordination structure and coordination parameters, it is able to find a solution — even for an unsafe set of queries, as long as all queries coordinate on the same parameters. This shows that by focusing on well structured and realistic scenarios rather than pathological ones, we can efficiently find coordination solutions to this otherwise NP-hard problem. In this algorithm as well, we do not guarantee to find the maximum size set that can coordinate; instead, we guarantee to find the maximum size coordinating set in which all queries get the same values for the coordination parameters.

The above algorithm provides a solution for coordination in most natural scenarios. In addition, we show that even the most minor extension where a subset of the users wants to coordinate on one additional attribute (for example the city from which a flight originates) is already not tractable.

**Organization of the Paper.** In Section 2, we formally define entangled queries and the problem of finding a coordinating set. Our contributions start in Section 3 where we cleanly characterize the sources of hardness in evaluating entangled queries. In Section 4, we show how to find a coordinating set if all the queries are safe, but not necessarily unique. In Section 5, we restrict our attention to applications in which users are coordinating with each other on the same set of attributes, and we give efficient algorithms for this case that no longer require either safety nor uniqueness. We present the results from a thorough experimental evaluation in Section 6. We have mentioned the most relevant related work in the introduction; for a more extensive discussion, we refer the reader to [5, 6].

## 2. PRELIMINARIES

### 2.1 Entangled Queries

We begin with a brief overview of entangled queries; for a more in-depth treatment, we refer the reader to the paper by

| $q$ | postconditions | head | body |
|---|---|---|---|
| $q_C$ | $R(G, x_1)$ | $R(C, x_1), Q(C, x_2)$ | $F(x_1, x), H(x_2, x)$ |
| $q_G$ | $R(C, y_1), Q(C, y_2)$ | $R(G, y_1), Q(G, y_2)$ | $F(y_1, P), H(y_2, P)$ |
| $q_J$ | $R(C, z_1), R(G, z_1)$ | $R(J, z_1), Q(J, z_2)$ | $F(z_1, A), H(z_2, A)$ |
| $q_W$ | $R(C, w_1), Q(J, w_2)$ | $R(W, w_1), Q(W, w_2)$ | $F(w_1, M), H(w_2, M)$ |

**Figure 1: Coordination Example**

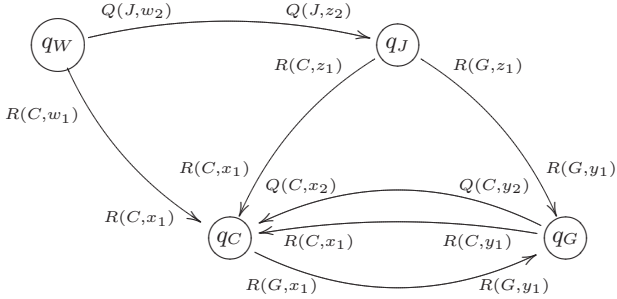

**Figure 2: Extended Coordination Graph**

Gupta et al. [5]. An *entangled query* is a triple $\{P\}\ H := B$, where $P$ is a list of *postcondition atoms*, $H$ is a list of *head atoms*, and $B$ is the *body* of the query, which is a list (conjunction) of atoms. The syntax of entangled queries also requires the following two properties: (i) All relation symbols that appear in the body are in the *database schema*. (ii) The relation symbols that appear in $P$ and $H$ are disjoint from those in the database schema; they are called *answer relation symbols*.

For example, suppose that Gwyneth wants to fly with Chris to Zurich, and that the database contains a relation F (flights) with attributes (flightId,destination). She might specify this with the following entangled query:

$$q_1 = \{R(\mathsf{Chris}, x)\}\ R(\mathsf{Gwyneth}, x) :-\ \mathsf{Flights}(x, \mathsf{Zurich})$$

**Definition 1 (Coordinating Set:)** *Let $I$ be an instance of a database, $Q$ a set of entangled queries, and $h$ be an assignment of values from the domain of $I$ to variables in $Q$. We refer to an atom where each occurrence of a variable $v$ is replaced by $h(v)$ as a grounded atom. We say that a non-empty subset $S \subseteq Q$ is a* coordinating set *if there exists an assignment $h$ of values from the domain of $I$ to variables in $S$ such that the following conditions hold:*
(1) *Every variable in $S$ is assigned a value under $h$.*
(2) *The grounded version of every body atom appears in $I$.*
(3) *The set of all grounded postcondition atoms that appear in $S$ is a subset of the set of grounded head atoms that appear in $S$.*

Suppose Chris issues the following query:

$$q_2 = \{\ \}\ R(\mathsf{Chris}, y) :-\ \mathsf{Flights}(y, \mathsf{Zurich})$$

Then the set $\{q_1, q_2\}$ is a coordinating set if there is a flight in the database to Zurich. For example, if the database contains the tuple $F(101, \text{'}Zurich\text{'})$, then the queries form a coordinating set under the assignment $h$ where $h(y) = 101$ and $h(x) = 101$.

The fundamental problem of evaluating entangled queries is given a set of entangled queries and a database instance, decide whether a coordinating set exists and – if so – find it. This is called a *choose-1* semantics [5], since only one answer to each query is desired, unlike in traditional conjunctive query answering. In our example, even if there are multiple flights to Zurich, it is understood that Gwyneth and Chris only want one flight number to be chosen and returned.

## 2.2 A Flight-Hotel Coordination Example

We now present a more elaborate coordination scenario which we later use to illustrate our algorithms.

Assume that Coldplay members Chris ($C$), Guy ($G$), Jonny ($J$) and Will ($W$) need a break from their tour and want to go on vacation. They want to book flights and hotels according to the following requirements:

- Chris wants to be on the same flight as Guy. He does not care about the destination.
- Guy wants to go to Paris and be on the same flight and hotel as Chris.
- Jonny wants to go to Athens and be on the same flight as Chris and Guy.
- Will wants to go to Madrid and be on the same flight as Chris and the same hotel as Jonny.

Assume the database contains the following relations: F (flights) with attributes (flightId,destination), and H (hotels) with attributes (hotelId,location). The band members will coordinate on the relations $R$ (flight coordination) and $Q$ (hotel coordination). The set of queries shown in Figure 1 expresses all the requirements stated above.

## 2.3 Coordination Graphs, Safety, and Uniqueness

Let $Q$ be a set of entangled queries. We define the *extended coordination graph* for $Q$ to be the directed multigraph with set of vertices equal to $Q$, and set of edges that contains all the pairs of the form $e = ((q, a_p), (q', a_h))$, where $a_p$ is a postcondition atom of query $q$, $a_h$ is a head atom of query $q'$, and $a_p$ *unifies* with $a_h$. We say that two atoms are unifiable if they are defined on the same relation and they do not contain different constants for the same attribute value. For example, $R(C, x_1)$ and $R(C, y_1)$ are unifiable whereas $R(C, x_1)$ and $R(G, y_1)$ are not.

For the set of queries $Q = \{q_C, q_G, q_J, q_W\}$ of the flights-hotels coordination example of Section 2.2, we have the extended coordination graph shown in Figure 2. The meaning of a directed edge $((q, a_p), (q', a_h))$ is that the postcondition atom $a_p$ of query $q$ *potentially needs* the head atom $a_h$ of $q'$ to be satisfied.

**Definition 2** *Safety: Let $Q$ be a set of entangled queries. We say that $q \in Q$ is* unsafe *in $Q$ if it has a postcondition atom that unifies with more than one head atom that appears in $Q$; otherwise query $q$ is* safe. *In other words, $q$ is unsafe in $Q$ if and only if there are at least two arrows emanating from $q$ with the same left-endpoint label in the extended coordination graph of $Q$. If all queries in $Q$ are safe then we say that the set of queries $Q$ is a* safe set.

The *coordination graph* of a set of queries $Q$ is constructed by collapsing parallel edges in the extended coordination graph. More formally, the set of its vertices is $Q$ and its edges are all the pairs $(q, q')$, where there is some postcondition atom in $q$ and some head atom in $q'$ that unify. For our example, we have the graph:

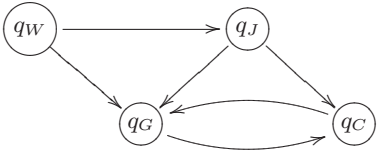

The coordination graph is useful for scenarios in which there is no ambiguity regarding the meaning of the edges. When the queries are safe, for example, an arrow from $q$ to $q'$ indicates that $q$ *needs* $q'$ to coordinate.

**Definition 3** *Uniqueness: We say that a set of safe queries $Q$ is* unique *if in $Q$'s coordination graph there exists a directed path between every two vertices.*

Note that uniqueness implies that the only way to satisfy the coordination requirements of one query in the set is to satisfy the coordination requirements of all of them. The entangled query evaluation algorithm presented by Gupta et al. finds a coordinating set if one exists for any set of queries that is both safe and unique [5]. The algorithm first checks whether all the queries can be unified together. This is done by traversing the queries in the order implied by the extended coordination graph and computing the Most General Unifier (i.e., the least restrictive unifier that enforces all the constraints imposed by the queries' head atoms). Then a combined query is constructed from the unified heads and bodies of all queries. This single query is then sent to the database and if successful it returns a valuation of the variables that witnesses the existence of a coordinating set.

## 3. HARDNESS OF ENTANGLED QUERIES

In this section, we present hardness results for entangled query evaluation. We show that entangled queries are associated with an extra source of hardness beyond the inherent hardness of evaluating conjunctive queries. Previous results have shown that one can encode NP-complete problems by using entangled queries with only one body atom [5]. However, these reductions relied on databases that were rich enough to allow the encoding of NP-complete problems in conjunctive queries by themselves. We give a crisp separation between the hardness of entangled queries and conjunctive queries by using instances where the database contains a single unary relation with two values and hence the satisfiability of every conjunctive query can be decided in polynomial time. We show that this is still enough to encode an instance of 3SAT with entangled queries. This implies that there is an additional source of hardness associated with finding a coordinating set of entangled queries. The reduction also allows us to pinpoint the specific source of hardness, namely, that finding a set of queries that can be *unified* is NP-complete.

We define the problem of entangled query evaluation more formally:

**Definition 4** ENTANGLED$(\mathcal{Q})$: *Given a set $Q$ of entangled queries in $\mathcal{Q}$ and a database instance $I$, does there exist a coordinating set?*

We define $\mathcal{Q}_{\text{all}}$ to be the class of all sets of entangled queries, and $\mathcal{Q}_{\text{safe}}$ to be the class of all safe sets of entangled queries. We now prove the following theorem:

**Theorem 1** ENTANGLED$(\mathcal{Q}_{\text{all}})$ *is NP-complete, even if the satisfiability of any conjunctive query on the database can be decided in polynomial time.*

PROOF OUTLINE. We reduce 3SAT to ENTANGLED$(\mathcal{Q}_{\text{all}})$. Consider an instance of 3SAT, i.e. a propositional formula that is represented as a set $\mathcal{C} = \{C_1, \ldots, C_k\}$ of clauses. The formula is the conjunction of the clauses and each clause is a disjunction of exactly three literals over the variables $x_1, x_2, \ldots, x_m$. We encode it as an instance $(Q, I)$ of ENTANGLED$(\mathcal{Q}_{\text{all}})$, where $Q$ is a set of queries and $I$ a database instance. We encode the satisfaction of the entire formula using the query

Clause-Query : $\{C_1(1), \ldots, C_k(1)\}\ C(1) :\!- \emptyset$

whose postconditions intuitively say that all clauses should be satisfied. A truth assignment to the variables is encoded as a selection of $R_i(1)$ or $R_i(0)$, corresponding to truth and falsity for the variable $x_i$ respectively. We use queries of the following form for this purpose:

$x_i$-Val : $\{C(1)\}\ R_i(x) :\!- D(x)$

$D$ is interpreted as a unary relation containing only the truth values 1 and 0. The purpose of the postcondition $C(1)$ is to create an interdependency among all the queries of the instance we define. It remains to give queries that describe how clauses are satisfied by the variable truth assignment.

$$x_i\text{-True}: \quad \{R_i(1)\} \bigwedge_{j:x_i \in C_j} C_j(1) :\!- \emptyset$$

$$x_i\text{-False}: \quad \{R_i(0)\} \bigwedge_{j:\neg x_i \in C_j} C_j(1) :\!- \emptyset$$

Since the database contains only a unary relation, any query to it is trivial, and hence can be implemented in polynomial time. It is clear that constructing $(Q, I)$ from $\mathcal{C}$ can be done in polynomial time.

If $x_i$ is true, then all clauses containing $x_i$ are satisfied. If $x_i$ is false, then all clauses containing $\neg x_i$ are satisfied. This completes our construction. In Appendix A we show that $\mathcal{C}$ is satisfiable iff the instance we have described is a "yes" instance of ENTANGLED. □

For instances in which there is more than one coordinating set we are faced with the problem of deciding which coordinating set the algorithm should return. A natural choice is returning a maximum-sized coordination set. This brings us to the following generalization of the problem ENTANGLED:

**Definition 5** ENTANGLEDMAX$(\mathcal{Q})$: *Given a set $Q$ of entangled queries in $\mathcal{Q}$, and a database instance $I$, find a maximum-sized coordinating set.*

We now show that even with the safety restriction, finding a maximum-sized coordinating set is NP-hard:

**Theorem 2** ENTANGLEDMAX$(\mathcal{Q}_{safe})$ *is NP-hard, even if the satisfiability of any query to the DB can be decided in polynomial time.*

PROOF OUTLINE. We give a reduction of 3SAT to ENTANGLEDMAX($Q_{safe}$). Let $\mathcal{C} = \{C_1, \ldots, C_k\}$ be an instance of 3SAT and $x_1, \ldots, x_m$ be the variables that appear in the instance. The selection of a truth value for $x_j$ is encoded as selection of $R_j(1)$ or $R_j(0)$ corresponding to truth and falsity respectively. This selection is forced by the query

$$q(x_j) = \{\} \, R_j(x_j) :\!- D(x_j) \;,$$

where $D$ is interpreted as a unary relation containing 0 and 1. If a clause $C_i$ contains a positive literal $x_j$, then the query

$$q(C_i, x_j) = \{R_j(1)\} \, C_i(1) :\!- \emptyset$$

says that a truth assignment that makes $x_j$ true satisfies $C_i$. If $C_i$ contained the negative literal $\neg x_j$ instead, we would introduce the query

$$q(C_i, \neg x_j) = \{R_j(0)\} \, C_i(1) :\!- \emptyset \;.$$

We are aiming for the claim: $\mathcal{C}$ is satisfiable iff the size of the largest coordinating set is $k+m$. However, the construction given so far cannot control the number of literals that satisfy a specific clause (more than one literal can satisfy a clause). This calls for the use of a more complicated gadget that ensures that the satisfaction of each clause is witnessed by only one literal. We leave the details for the appendix. □

Given the above hardness result, the algorithms we present in subsequent sections use criteria other than maximality to decide which coordinating set to return.

## 4. LIFTING UNIQUENESS

We are now ready to present the algorithm for finding a coordinating set when the queries are safe. Recall that the algorithm by Gupta et al. [6] was able to find a coordinating set only for a set of queries which is both safe and unique. Here we show that the uniqueness property is not required for making the problem tractable. We begin by describing our general approach, and then present our algorithm.

Let $Q$ be a safe set of queries. The main observation is that due to safety, if $q$ belongs to a coordinating set $S$ then all its successors in the coordination graph $G$ have to be in $S$ as well. This implies that every strongly connected component (SCC) of $G$ is either contained in $S$ or disjoint from $S$, because in such a component all queries depend on one another. This observation suggests that we can contract each SCC of $G$ into a single query and proceed to solve a problem with a simpler coordination graph: the resulting graph is a directed acyclic graph (DAG). We call this graph the *components graph of $G$* and denote it by $G'$. More formally, $G'$ has the SCCs of $G$ as vertices and we put an edge from a SCC $S_1$ to a SCC $S_2$ if there are vertices $u \in S_1$, $v \in S_2$ such that $(u,v)$ is an edge of $G$.

Consider the example from Section 2.2. The SCCs of the coordination graph are $\{q_C, q_G\}$, $\{q_J\}$, $\{q_W\}$. We will use this example in our explanations below.

Once we have constructed the components graph, we process the components in reverse topological order. At every step of the algorithm we process one vertex of $G'$, which corresponds to one SCC of $G$. We unify the queries of this component with the combined queries of its successor components. The resulting query is then sent to the database for evaluation. Processing in reverse topological order ensures that in every step of the algorithm, all of the successors of this component in $G'$ have already been processed. If one of the successors fails, we can stop processing the current component as well.

In our example graph, the first node of the components graph $G'$ that we analyze is $\{q_C, q_G\}$ as it has no outgoing edges. We perform all the necessary unifications to obtain a single query with no postconditions. The queries $q_C$, $q_G$ after the unifications give us the conjunctive query $(q_C + q_G)$:

$$\{\ \} \quad R(C, x_1), Q(C, x_2), R(G, x_1), Q(G, x_2) :\!-$$
$$\text{Flights}(x_1, x), \text{Hotels}(x_2, x),$$
$$\text{Flights}(y_1, \text{Paris}), \text{Hotels}(y_2, \text{Paris})$$

Next, the combined query $(q_C + q_G)$ is issued to the database. If it is unsuccessful, then the algorithm can declare that there is no coordinating set. This is because all other SCCs depend on this one. Otherwise, the queries $q_C$ and $q_G$ are a coordinating set and Chris and Guy can travel together. The algorithm can now continue and try to find a larger coordinating set by collapsing the SCC $\{q_C, q_G\}$ and considering the new graph:

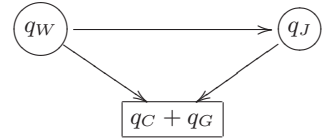

The algorithm can unify $q_J$ with $q_C + q_G$ to obtain $(q_C + q_G) + q_J$ (if unifications succeed) and send the query to the database. If successful, it attempts to unify also with $q_W$, and if successful send it also to the database.

The number of queries that the algorithm sends to the database is bounded by the number of SCCs of the reduced coordination graph and hence bounded by the size of $Q$. Also, note that by going over all the SCCs in reverse topological order our algorithm does not duplicate work. The extra work needed besides issuing the query is at most quadratic in the number of queries. All we need to do is build a graph (in the worst case the number of edges is quadratic in the number of queries, although usually the graph will be very sparse), find its strongly connected components and traverse them in reverse topological order.

When the algorithm finishes, we might have a list of coordinating sets instead of only one. We use the most natural way to choose between them, namely take a maximal one. For example, consider the following components graph:

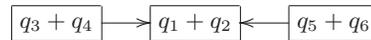

Assuming that all unifications and queries to the database succeed, the algorithm discovers the coordinating sets

$$\{q_1, q_2\}, \{q_1, q_2, q_3, q_4\}, \{q_1, q_2, q_5, q_6\} \;.$$

The last two sets in the above list have maximum size, and the algorithm chooses one of them arbitrarily. Different applications might want to consider other selection criteria. For example, an airline website might choose to include the set with the most number of passengers with gold status, or taking the set containing some specific query (if such exists) representing some VIP client. These are simple changes that can be made to the last step of the algorithm.

One should notice that the algorithm does not compute all possible coordinating sets. For the example of the previous paragraph, the algorithm does not check whether

$$\{q_1, q_2, q_3, q_4, q_5, q_6\}$$

is a coordinating set, even though it is possible that it is. So, our algorithm is not guaranteed to find a coordinating set of maximum size (among all the possible ones); however, this guarantee is impossible to achieve without sacrificing tractability due to Theorem 2. Our algorithm does provide a slightly weaker guarantee. To describe it, we introduce one extra piece of notation: let $R(q)$ be the set of queries $q'$ such that $q'$ is in a SCC reachable from the SCC $q$ belongs to. For the example of the previous paragraph:

$$R(q_1) = R(q_2) = \{q_1, q_2\}$$
$$R(q_3) = R(q_4) = \{q_1, q_2, q_3, q_4\}$$
$$R(q_5) = R(q_6) = \{q_1, q_2, q_5, q_6\}$$

Our algorithm is guaranteed to choose a coordinating set of maximum size among the coordinating sets that belong to $\{R(q) \mid q \in Q\}$.

We now present the full SCC coordination algorithm.

*The SCC Coordination Algorithm*
- Construct the coordination graph $G$ and partition it into strongly connected components (SCCs), to get the components graph $G'$.
- Go over the nodes of $G'$ in reverse topological order and for each node:
  – Check that none of its successors has failed to coordinate. If some successor has failed, then the current node is also marked as failed and we continue to next node.
  – Unify all the queries that belong to the SCC in $G$ that corresponds to the current node together with the combined queries of the successor nodes. If unification fails, mark the current node as failed and continue to next node.
  – Create the combined query and try to ground it by querying the database. If successful, store the combined query for the node, the total number of original entangled queries which are involved, and the computed grounding.
- Take the set that represents the largest number of queries and return it as the coordination set.

**Running Time.**
- Creating the components graph requires time linear in the size of $G$.
- Every query is only unified once with its successors, so the total time spent on unifications is linear in the size of $Q$.
- The traversal of $G'$ in reverse topological order can be done in time linear in the size of $G'$.
- At most $|Q|$ queries are issued to the database.

Thus, in total we have $|Q|$ queries to the database and extra processing overhead which is quadratic in $|Q|$.

## 5. BEYOND SAFETY

In the previous section we saw that the lack of uniqueness is not a real obstacle for finding a coordinating set. However, the safety property does seem crucial in making the general problem tractable. It is possible to formulate restrictions other than safety that can allow tractable solution in the general case. One example is the following.

**Definition 6 (Single-connectedness)** *Let $Q$ be a set of entangled queries; it is* single-connected *if the following two properties hold. First, all queries have at most one post-condition atom. Second, for every pair $q$, $q'$ of queries in $Q$, there is at most one simple path (no repetition of edges) from $q$ to $q'$ in the coordination graph for $Q$. We denote this class of sets of queries as $\mathcal{Q}_{sc}$.*

**Theorem 3** ENTANGLED2$(\mathcal{Q}_{sc})$ *can be solved with a linear number of conjunctive queries (of linear size) to the database.*

However, it is unclear whether single-connectedness is a natural property typically found in real-world coordination. Therefore, rather than look for further such properties, we take a different approach in this section. We suggest a class of application-specific algorithms that take advantage of knowledge of the particular characteristics of an application in order to find a coordinating set, even for an unsafe set of queries. While our approach is application-specific, the constraints we impose are relatively mild and are likely to apply in most realistic coordination scenarios.

All of the coordination examples we have seen so far, such as deciding on a flight, course or meeting time together, share a common feature: the users want to coordinate on some common attribute(s) - for example, fly to the same destination on the same day, or go to the same party. We refer to these common attributes as coordinating attributes. The property that all users are coordinating on the same attributes is quite powerful since it assures us that if there exists a coordinating set, there is a coordinating set in which all users "agree" on the coordination attributes. We refer to queries that satisfy this property as *Consistent Queries* and define them formally next.

Our overall approach is the following. By knowing in advance what the possible coordination attributes are, and in addition taking advantage of knowledge of the social relations between the users, we can just take from the database the possible values for these attributes, find for each value the set of queries that are satisfied using it, and then check that their coordination requirements hold.

*Consistent Queries*

We now describe more precisely the application-specific restrictions and assumptions we place on entangled queries. Our first restriction is on the users' relationships. We consider a group of users who want to coordinate and assume each one submits *a single* query. The social relationships among the users are given by a single binary relation, named F, in the database. For simplicity, we assume that the database contains a single additional relation named S with attributes $A_1, \ldots, A_d$. For example, for the flights reservation scenario, this can be the relation containing the details of the flights.

Our second restriction is on the queries' structure. Let User be the name of a user who submits a query to the system. The general form of his query is the following:

$$\{R(y_1, f_1), R(y_2, c_2), \ldots, R(y_k, c_k)\}$$
$$R(x, \mathsf{User}) :\!\!- \mathsf{S}(x, a_1^x, \ldots, a_d^x), \mathsf{F}(\mathsf{User}, f_1),$$
$$\bigwedge_{i=1}^{k} \mathsf{S}(y_i, a_1^i, \ldots, a_d^i)$$

In the above query, user User wants to coordinate with $k$ users $f_1, c_2, \ldots, c_k$. $f_1$ is a variable that intuitively represents a friend of User in the relation F, i.e. $\mathsf{F}(\mathsf{User}, f_1)$.

The rest of the coordination partners are given by constants $c_2, \ldots, c_k$. The answer relation $R$ consists of (key[1], username) pairs. Such pairs carry all the relevant information, since the key uniquely identifies the tuple satisfying the user's requirements. So, $x$ is the key for the tuple satisfying User's requirements, and $y_1, \ldots, y_k$ are the keys for the tuples satisfying his friends' requirements. The terms $a_j^x$ ($j = 1, \ldots, d$) and $a_j^i$ ($i = 1, \ldots, k, j = 1, \ldots, d$) can be either constants or variables.

A user wishing his coordination partner $f_i$ to receive the same tuple as him (e.g., the same flight) can define $y_i = x$.

As previously discussed, we would like to restrict our attention to set of queries such that all users are coordinating on the same attributes. To make this precise, we will define the notions of $A$-coordinating, $A$-non-coordinating, $A$-consistent where $A$ is a subset of the attributes of the relation S.

**Definition 7 ($A$-coordinating)** *A query $q$ is $A$-coordinating if for every attribute $A_j \in A$, User specified the same constants or variables for himself and for all his coordination partners. More formally, for every $A_j \in A$ and $i = 1, \ldots k$ it is the case that $a_j^x = a_j^i$.*

**Definition 8 ($A$-non-coordinating)** *A query $q$ is $A$-non-coordinating if for every attribute $A_j \in A$, User may specify a constant for himself but not for his coordination partners. More formally, for every $A_j \in A$ and $i = 1, \ldots k$, all $a_j^i$'s are distinct variables. In case $a_j^x$ is a variable it is also distinct.*

We note that the notion of $A$-non-coordinating is *not* the negation of $A$-coordinating. Observe that a query is not $A$-coordinating if there is an attribute $A_j$ in $A$ such that $a_j^x, a_j^1, \ldots, a_j^k$ are not all the same. The notion of $A$-non-coordinating implies something stronger, namely that for every $A_j$ in $A$ the $a_j^1, \ldots, a_j^k$ are all distinct.

**Definition 9 ($A$-consistent)** *A query $q$ for a table $S$ where $\mathcal{A}$ is the set of attributes is $A$-consistent if it is $A$-coordinating and $(\mathcal{A} - A)$-non-coordinating.*

The following proposition follows from the definition:

**Proposition 1** *Let $Q$ be a set of queries and $A$ a set of attributes where every $q$ in $Q$ is $A$-consistent. Then, there exists a coordinating set if and only if there exists a coordinating set with all tuples agreeing on the attributes in $A$.*

### Algorithm by Example

Since our previous example was safe we present a new example to illustrate the idea behind the algorithm. This time, the members of Coldplay have a night off and want to catch a movie. Everyone wants to save on cab fare, so each band member wants to go to a cinema with at least one other band member. They do not care whether they go to the same movie with the other band member or not. They do, however, have preferences regarding the movie and/or the cinema. Denote by $C$ the table describing the friendships between Coldplay members which contains the following information:

[1] For simplicity we assume that the relation S has one column which is a unique key. However, all definitions can be generalized for the case that the key is combination of attributes.

- Chris's friends are: Jonny, Guy.
- Guy's friends are: Chris and Jonny.
- Jonny's friends are: Chris and Will.
- Will's friends are: Chris and Guy.

Denote by M=(movie_id, cinema_name,movie_name) the cinemas table where movie_id is the key, and consider the following set of queries:

$$q_c = \{R(y, \mathsf{Will})\} R(x, \mathsf{Chris}) :-$$
$$M(x, \mathsf{Regal}, \mathsf{Contagion}), M(y, \mathsf{Regal}, z)$$
$$q_g = \{R(y, f)\} R(x, \mathsf{Guy}) :-$$
$$C(f, \mathsf{Guy}), M(x, \mathsf{AMC}, \mathsf{Project\ X}), M(y, \mathsf{AMC}, z)$$
$$q_j = \{R(y, f)\} R(x, \mathsf{Jonny}) :-$$
$$C(f, \mathsf{Jonny}), M(x, b, \mathsf{Hugo}), M(y, b, z)$$
$$q_w = \{R(y, f)\} R(x, \mathsf{Will}) :-$$
$$C(f, \mathsf{Will}), M(x, b, \mathsf{Hugo}), M(y, b, z)$$

Notice for the $q_c$ query that Will is not a friend of Chris, yet it is possible for Chris to submit a query where the constant Will appears in the postconditions.

We begin by computing for each query $q$ the list of values for the coordination attributes that satisfy the query requirements. We denote this list by $V(q)$. More Formally:

**Definition 10** *Let $q$ be an $A$-consistent query. We define $V(q)$ to contain all tuples $v = (v_k)_{k \in A}$ of values indexed by attributes in $A$, such that:*
*(i) If $q$ specifies a constant for the attribute $k \in A$, then $v_k$ is equal to that constant.*
*(ii) If we substitute for every attribute $k \in A$ the value $v_k$ in the body of the query $q$, then the body of the query is satisfiable.*

For the movies example, to construct $V(q_j)$ and $V(q_w)$ we should know where the movie Hugo plays. Assume it plays at Regal, AMC and Cinemark. In the table below we present the options list for each of the band members.

| Band Member | Possible cinemas |
|---|---|
| Chris | $V(q_c) = \{$ Regal $\}$ |
| Guy | $V(q_g) = \{$ AMC $\}$ |
| Jonny | $V(q_j) = \{$ Regal, AMC ,Cinemark $\}$ |
| Will | $V(q_w) = \{$ Regal, AMC ,Cinemark $\}$ |

Next, we construct the *pruned coordination graph*. This graph is a subgraph of the coordination graph that contains only queries whose bodies can be satisfied. Moreover, an edge from a query $q$ to a query $q'$ is kept only if the unification of the queries does not cause an inconsistency with the "friendship" relation. We describe the pruned coordination graph for the example above. We already know that for each band member there exists a tuple satisfying his query's body. As for the edges, the edge $(q_i, q_j)$ exists if either $i$ specifically specifies that he wants to coordinate with $j$ (for example the edge from Chris to Will) or that $j$ is a friend of $i$ according to table $C$. For example, Guy is not a friend of Jonny and hence the edge from Jonny to Guy is not part of the pruned coordination graph. The coordination graph and pruned coordination graph for the example are depicted in Figure 3.

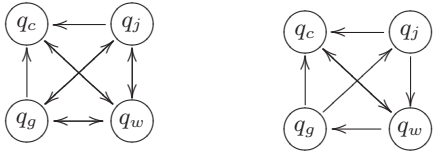

**Figure 3: Coordination Graphs**
On the left, the coordination graph for the movies example and on the right, the pruned coordination graph.

In the next step, we create a list of optional values for the coordination attributes of a coordinating set if one exists. We denote this list by $V(Q)$. This is simply the union of $V(q)$ for all $q \in Q$. In our example $V(Q) = \{\mathsf{Regal}, \mathsf{AMC}, \mathsf{Cinemark}\}$.

Now, to find a coordinating set (if one exists) the algorithm can simply go over the values in $V(Q)$, for each $v \in V(Q)$ it considers the subgraph of the pruned coordination set restricted to queries $q_i$ such that $v \in V(q_i)$ which we denote by $G_v$. Then it does a cleaning step by removing from $G_v$ queries whose coordination requirements don't hold. A query's coordination requirements don't hold if either there is no edge between it and one of its specified coordination partners (was specified by a constant) or it specifies a friend from the table as one of its coordination partners and there are no edges between it and any of its friends. If at the end of the process $G_v$ is not empty then it has found a coordinating set. It can keep going over the values in $V(Q)$ till it finds the one for which the coordinating set is maximal, or until another appropriate criterion for selecting a coordinating set is satisfied.

In our example, consider $G_{\mathsf{Cinemark}}$. The only members that can go to Cinemark are Jonny and Will, hence $G_{\mathsf{Cinemark}}$ only contains both of them. It is easy to see that the coordination requirement of Will is not satisfied since he wants to go to the cinema with one of his friends, and $G_{\mathsf{Cinemark}}$ contains none of his friends. Thus, we remove $q_w$ from the graph. Now, Jonny's coordination requirements are also unsatisfied hence we remove him as well. We conclude that there is no coordinating set that can go to Cinemark. Now, consider $G_{\mathsf{Regal}}$. The only member that cannot go to Regal is Guy, hence $G_{\mathsf{Regal}}$ contains all members except for Guy. It is not hard to verify that in $G_{\mathsf{Regal}}$ the coordination requirements of Chris, Jonny, and Will are fulfilled and hence this is a coordinating set.

The algorithm can be specified more formally as follows:

*The Consistent Coordination Algorithm*

Let $Q$ be a set of $A$-consistent queries.
- For every query $q \in Q$, create $V(q)$ (the list of all values for the coordination attributes that satisfy $q$).
- Construct the pruned coordination graph over all queries $q$ such that $V(q)$ is non empty.
- Let the options list $V(Q) = \bigcup_{q \in Q} V(q)$.
- For every $v \in V(Q)$:
  - Create the subgraph $G_v$ of all queries for which $v \in V(q)$.
  - Execute a cleaning phase: iteratively remove from $G_v$ all queries which have a postcondition that cannot be satisfied.
  - If $G_v$ is nonempty record its size.

The algorithm can find several coordinating sets and uses an appropriate criterion such as maximality to choose which one to return.

**Running time**. The running time of the algorithm can be described in terms of the following:
- Computing $V(Q)$ and $v(q)$ $\forall q \in Q$ requires $O(n)$ queries to the database.
- The construction of the pruned coordination graph requires time $O(n^2)$.
- Let $M$ be the size of $V(Q)$. The last step of the algorithm performs $M$ iterations, each one has a cleaning phase which take $O(n^2)$.

This means that the algorithm is linear in the size of the database and quadratic in the number of users.

*Discussion*

**Limits**. We have exhibited a fragment of entangled queries for which there exists an efficient algorithm that finds a coordinating set if one exists. It is natural to ask whether this fragment can be generalized without sacrificing tractability. We show in Section B of the Appendix that by allowing some queries to coordinate on some attribute $a$ and some to coordinate on attributes $a$ and $b$ we can encode 3SAT and therefore the problem becomes NP-hard.

**Generalizations**. We have described our idea in its simplest form for clarity of exposition; however, it lends itself to natural extensions. The coordination attributes do not need to come from only one relation, as long as all users coordinate on all of them.

By trivially tweaking the graph construction step and the cleaning step, the algorithm can also handle queries which use more than one binary relation to specify coordination partners. The algorithm now needs to check more conditions when deciding when to remove a query from the graph. Similarly, the cleaning procedure can be adapted to allow users to specify that they want to coordinate with $k$ friends from some table. Interestingly, this latter type of coordination is not even expressible in the current entangled query syntax.

## 6. EXPERIMENTAL EVALUATION

We now give implementation details, experimental setup and results for both our algorithms.

### 6.1 SCC Coordination Algorithm

*Implementation*

We implement the SCC Coordination Algorithm presented in Section 4 as part of the *Youtopia* system [5]. Our implementation extends the query evaluation process in the coordination module of the system. The system works roughly as follows: when a new query arrives, the system finds the set of queries this query can coordinate with and updates the coordination graph accordingly. The system then calls an evaluation method on the connected component that the query belongs to.

In the original algorithm of Gupta et al. [5], the evaluation method computes a combined query, which aggregates the requirements of all the queries in the connected component into a single query, and then issues it to the database. In the SCC Coordination Algorithm, on the other hand, it first preprocesses the connected component and iteratively remove all queries that have an unsatisfiable postcondition. The algorithm then finds the strongly connected components, and

1179

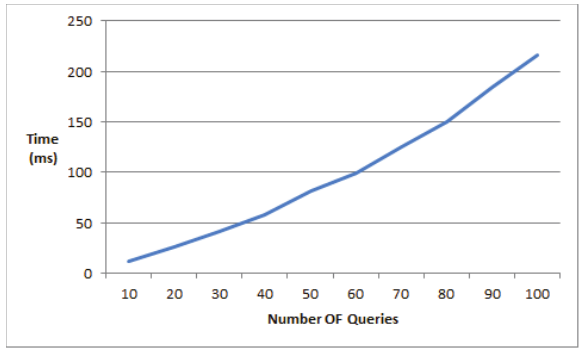

Figure 4: Processing Time in List Structure

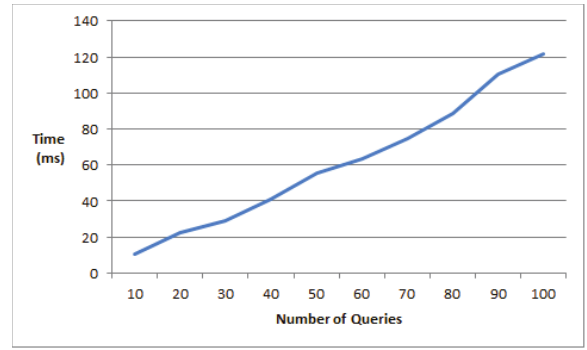

Figure 5: Processing Time in Scale Free Network Structure

uses a Hash table to map each query to its strongly connected component. Next, it builds a new graph, where each node represents a strongly connected component. An edge in the new graph exists if there is an edge between queries belonging to both components in the coordination graph.

Next, the algorithm processes the graph in reverse topological order. For each node, it unifies the queries corresponding to that node with the combined queries that resulted from its successors. If one of these unifications fail, it stops processing this node and continues to the next node.

For every successful unification the algorithm then tries to ground the resulting combined query. If there is a tuple satisfying this query, it stores in a Hash table the combined query for this set and the set of queries it included, to be used by other nodes that are connected to it. If the size of this set is larger than the largest coordinating set found so far, the new set is stored as the maximum sized set. Finally, the algorithm returns the largest successful query set. The system then deletes these queries from its data structures and continues to process the next query that arrives.

The implementation is in Java. We use the Java package JGraphT for graph operations [8]. We use a MySQL database, and JDBC for submitting queries to the database.

*Experiments*

To evaluate the algorithm, we use a scenario where users post queries looking for other specific users to coordinate with. As in [5] we use the *Slashdot* social network data. The table we query has 82168 entries. The bodies of the entangled queries are simple and we make sure that for each body there is at least one tuple satisfying it in the database. This is the most demanding scenario for finding a coordinating set, as when a query's body cannot be satisfied, the algorithm simply removes its strongly connected component and all components which depend on it. This reduces the complexity of the resulting coordination.

Our main goals in these experiments are measuring the total running time of the algorithm, measuring the overhead caused by building the graph and running the graph algorithms, and examining scalability with respect to the number of queries. We performed all experiments on an Intel i3 CPU M350 @2.27GHz with 4GB of RAM.

We start with a simple structure involving a list of queries where each query requests to coordinate with the next query on the list, while the last query does not require any coordination partner. This yields a non-unique coordination struc-ture which cannot be handled by previous algorithms. We choose this structure since it is the worst case for our algorithm – there is a different coordinating set for every query we consider, and therefore the largest number of database queries are issued. We run this experiment for various list sizes. It is clear from Figure 4 that the processing time grows linearly with the number of queries. Our upper bound of 100 queries is very generous for a real-world set of connected queries that all want to coordinate with each other.

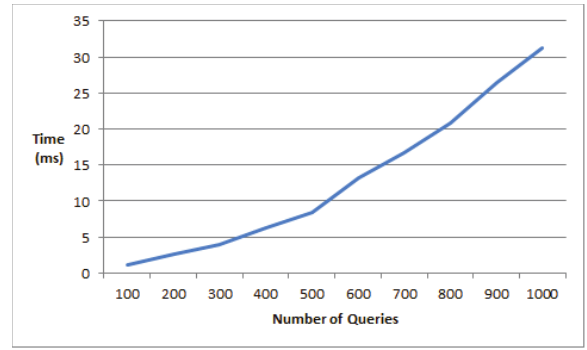

Figure 6: Graph Processing Time in Scale Free Network

For our next experiment we use a scale free directed network to decide on the coordination requirements of each query [1]. The degree distribution of nodes in these networks follows a power law: only a small number of nodes have high in-degree and most of the nodes have small in-degrees; these are considered a reasonable model of social networks. Each query in our workload corresponds to a node in the graph and its coordination partners are its successors in the graph. Clearly, this set of queries is not unique either. We examine the performance for sets of queries of varying sizes. Since the construction of the graph is a random process we average the running time over ten different randomly generated graphs of the same size. The results are shown in Figure 6.1. It is easy to see that the running times are now shorter, and that the growth remains linear in the size of the set.

In our last experiment we stress test the graph construction and preprocessing. We again use scale free networks, this time of sizes 100 to 1000 and look at the time needed for the graph construction and preprocessing. For each op-tional size we again use ten different randomly generated graphs. As Figure 6 shows, even for very large coordination graphs, the graph processing time is negligible, and grows very slowly.

## 6.2 Consistent Coordination Algorithm

*Implementation*

The *Youtopia* System provides a very general solution for entangled queries that works for any database schema and any query structure as long as the set of queries is safe; the uniqueness requirement was dropped by our previous algorithm. Recall that our approach for answering coordination queries that are unsafe is application driven and hence very different. It needs to be tailored to the schema of the application (or family of applications) and needs to know specific details about the coordination characteristics — this requirement is essentially what allows our approach to solve the coordination problem for unsafe queries efficiently. This observation is what led us to decide to implement the Consistent Coordination Algorithm as a separate prototype and not as a part of the *Youtopia* System. Due to this choice we cannot compare the running times of both algorithms. However, since the two algorithms are designed to solve almost completely different problems, comparisons between their running time would not be meaningful.

Our prototype uses the flight coordination example as our driving application scenario. This scenario lets each user coordinate with one of their friends on flying to the same destination on the same day. In addition each user can specify the source airport and airline they want to use. In other words, the coordination attributes are day and destination, while the non-coordination attributes are source and airline. For each attribute the user can specify a constant, or a special "don't care" value, meaning there is no restrictions on the value of this attribute. In addition each user specifies a friend they want to coordinate with, or specifies a "don't care" value, indicating she wants to coordinate with any of her friends that are defined in a predefined Friends table. The algorithm takes as input a set of queries in the form described; these are buffered and processed in batches, as done in existing entangled query processing algorithms.

The algorithm uses three types of database queries. The first type is used to retrieve all the values that satisfy each entangled query's coordination attributes. The second type of query is used to retrieve the list of friends of each entangled query from the predefined friends table. The friends list of each query is subsequently used to build the pruned coordination graph. The last type of query is issued after the largest set is found by the algorithm. The body of each entangled query in the set is sent to the database with the resulted values of the coordination attributes to get a specific flight number matching its body. In total, the number of database queries that the algorithm issues is linear in the number of entangled queries.

The memory requirements of the algorithm are small, since the size of the graph is at most quadratic in the number of entangled queries. In most reasonable applications the graph will be very sparse, hence close to linear size. In addition, for each query we keep a list of possible values for the coordination parameters that together with its other parameters can be satisfied by a flight in the database. In the worst case, the number of values stored for each query is the number of distinct coordination attributes values that exists in the database. Therefore, the amount of memory used is a function of the number of queries times the size of the DB table. However, in most reasonable scenarios there is only a small constant number of possibilities that each user considers; for example, a user is unlikely to consider more than 100 possible flights for a trip. Given this assumption, the memory needed is linear in the number of queries.

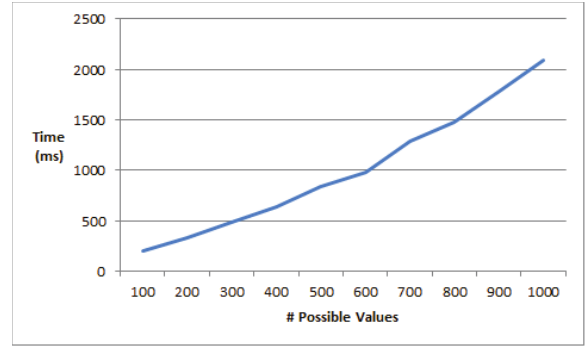

**Figure 7: Processing Time as a Function of Possible Values**

The algorithm's final output is a mapping between the users in the coordinating set to their selected flight number.

We implement the prototype in Java. The graph is built and maintained throughout the algorithm's run using the JGraphT package. We use a MySQL database and JDBC for submitting queries to the database.

*Experiments*

Our goal in these experiments is to stress test the algorithm, and show that it scales nicely and performs well even with a very large and thus unrealistic number of queries and attributes values. We used the same Intel i3 CPU M350 @2.27GHz with 4GB of RAM for this set of experiments.

For the first experiment we use a fixed set of 50 queries and vary the size of the Flights Table in the database. The Friends table encodes a complete friendship graph between all the users. In addition, the queries are such that all the values in the DB satisfy them. Notice that this is the absolutely worst possible scenario, since for every possible value of the coordination attributes, no queries are pruned in any step of the algorithm. Also, the number of possible values for the coordination attributes that the algorithm considers equals the number of distinct values for the coordination attributes in the database. For these experiments we used Flights tables with sizes varying from 100 to 1000. All flights in the table are unique, hence the number of options for the coordination attributes grows with the size of the table.

Note that this scenario has a large number of queries and each query has a huge number of possible values. As argued before, we believe that this is not a realistic scenario and that most expected scenarios exhibit much fewer queries and values, but it is a test for showing how are our algorithm scales. As can be seen in Figure 7, the processing time grows linearly with the number of options for the coordination attributes.

Our second experiment focuses on a more realistic scenario where the number of possible options is much smaller. We achieve that by using a Flights table that consists of only 100 tuples, each tuple holds a different destination and date combination. We use sets of queries of size 10-100. As in the previous experiment, the Friends table encodes a complete friendship graph. All the queries are such that every tuple in the DB satisfies them, which is the worst case for our algorithm. Our main goal is to show scalability with the number of queries. As can be seen in Figure 8, the processing time indeed grows linearly with the number of queries.

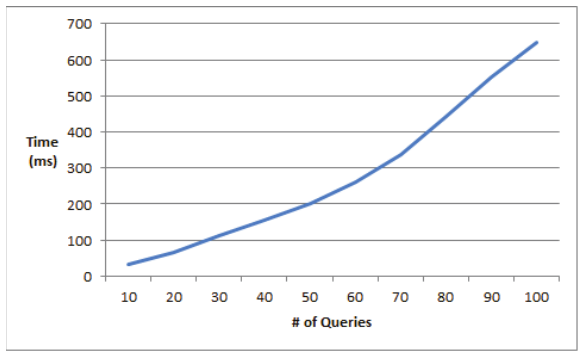

**Figure 8: Processing Time as a Function of Number of Queries**

We decided to focus on these two worst case examples, as the performance of the algorithm is highly dependent on the structure of the queries and the database. We believe that running the algorithm on other instances cannot provide any further insights on the behavior of the algorithm, since the results will be specific to the instance in question. On the other hand, the trends of these two experiments easily show, that in a more realistic setting with a more restricted coordination instance, the algorithm will perform very well.

A last point worth mentioning is that our implementation does not use any parallelism, although our algorithm naturally breaks into parallel processes, where each possible value can be easily checked independently. We believe that this could even further reduce the running time, but we leave this enhancement open for future work.

## 7. FUTURE WORK

Our results significantly expand the range of scenarios for using entangled queries and set the ground for the implementation of real applications. The obvious next step is choosing one or more real social coordination application and implementing it using our algorithms, as part of a full scale system for declarative, data-driven coordination as suggested by Gupta et al. [6].

Our work also raises additional interesting research questions. How can our algorithms be modified for an on-line setting where users continuously submit new queries? In particular, can the coordinating set(s) be found efficiently in an incremental fashion? Also, can we design internal database structures to support and optimize the execution of our algorithms, and perhaps even adapt to workload changes in the spirit of automatic database tuning? Finally, it will be interesting to compare our results with those in emerging new coordination formalisms such as Enmeshed Queries [2].

## 8. ACKNOWLEDGMENTS

This research has been supported in part by the NSF under Grants IIS-0910664, CCF-0910940, IIS-0911036, IIS-0812045, and IIS-1012593, by AFOSR grants FA9550-08-1-0438 and FA9550-09-1-0266, by ARO grant W911NF-09-1-0281, by the iAd Project funded by the Research Council of Norway, and by a Google Research Award. Any opinions, findings, conclusions or recommendations expressed in this paper are those of the authors and do not necessarily reflect the views of the sponsors.

## 9. REFERENCES

[1] A. Barabási and R. Albert. Emergence of scaling in random networks. *Science*, 286(5439):509, 1999.
[2] J. Chen, A. Machanavajjhala, and G. Varghese. Scalable social coordination using enmeshed queries. http://arxiv.org/abs/1205.0435, May 2012.
[3] S. Dalal, S. Temel, M. Little, M. Potts, and J. Webber. Coordinating business transactions on the web. *IEEE Internet Computing*, 7(1):30–39, 2003.
[4] H. Garcia-Molina and K. Salem. Sagas. *ACM SIGMOD Record*, 16(3):249–259, 1987.
[5] N. Gupta, L. Kot, S. Roy, G. Bender, J. Gehrke, and C. Koch. Entangled queries: enabling declarative data-driven coordination. In *ACM SIGMOD*, pages 673–684, 2011.
[6] N. Gupta, M. Nikolic, S. Roy, G. Bender, L. Kot, J. Gehrke, and C. Koch. Entangled transactions. *PVLDB*, 4(11):887–898, 2011.
[7] http://hatchplans.com.
[8] http://www.jgrapht.org/.
[9] http://www.trooptrip.com.
[10] L. Kot, N. Gupta, S. Roy, J. Gehrke, and C. Koch. Beyond isolation: Research opportunities in declarative data-driven coordination. *ACM SIGMOD Record*, 39(1):27–32, 2010.
[11] J. R. Larus and R. Rajwar. *Transactional Memory*. Morgan and Claypool, 2007.
[12] R. Milner. *Communicating and Mobile Systems: the Pi-Calculus*. Cambridge University Press, 1999.
[13] J. Roberts and K. Srinivasan. The tentative hold protocol. W3C Note, www.w3.org/TR/tenthold-1/., Nov 2001.
[14] G. Weiss. *Multiagent systems: a modern approach to distributed artificial intelligence*. MIT Press, 1999.

## APPENDIX
## A. PROOFS

**Proof of Theorem 1.** $\mathcal{C}$ is satisfiable iff the instance we have described is a "yes" instance of ENTANGLED.

- $\mathcal{C} \in 3\text{SAT} \implies (Q, I) \in \text{ENTANGLED}$: Let $h$ be a variable assignment that satisfies $\mathcal{C}$. This assignment $h$ tells us which queries to choose for the coordinating set $S \subseteq Q$. First, we put Clause-Query in $S$. For every variable $x$, we put the query $x$-Val in $S$ and also exactly one of the queries $x$-True, $x$-False: we put the query $x$-True ($x$-False) if $h(x) = \text{True}$ ($h(x) = \text{False}$). All the postconditions of the variable queries are satisfied since we only take exactly one of the $x$-True and $x$-False queries for every $x$, the interpretation of $D$ in $I$ contains 0, 1, and Clause-Query is in $S$. As for satisfying the Clause-Query, since $h$ is a satisfying assignment, for each clause $C_j$ there is at least one variable $x_i$ satisfying it. When we construct the coordinating set we include in it the query associated with this variable. Therefore for every clause $C_j$ the post condition of Clause-Query associated with it is satisfied.

- $(Q, I) \in$ ENTANGLED $\implies \mathcal{C} \in$ 3SAT: Let the coordination set of queries in $Q$ be $S$. Any such set $S$ must include the Clause-Query query, otherwise, it cannot include any $x_i$-Val query which implies it cannot include any $x_i$ query at all, and thus it is empty. Therefore it includes the Clause-Query. This query requires that all of its postconditions are satisfied. This means that for each postcondition $C_j(1)$ there is at least one query in $S$, for which $C_j(1)$ is one of its head atoms. This can be either an $x_i$-True query or a $x_i$-False query for some $i$. We construct a satisfying assignment $h$ for $\mathcal{C}$, by setting $x_i$ to True if the $x_i$-True query is in $S$ and to False if the $x_i$-False query is in $S$. Notice that for each $x_i$ it is impossible for both $x_i$-True and $x_i$-False to be in $S$ as they cannot both be unified with the val-query. If neither queries are in $S$ we can arbitrarily set it to True. Since each postcondition of the Clause-Query is satisfied we have that each clause in $\mathcal{C}$ is satisfied by $h$, hence $\mathcal{C} \in$ 3SAT.

∎

**Proof of Theorem 2.** Notice that the collection of all the $q(C_i, x_j)$ and $q(x_j)$ queries is not good enough to encode the 3SAT instance. This is because, we cannot control the number of $q(C_i, x_j)$ queries that will be satisfied: A clause can be satisfied through more than one literal.

We will use a gadget that guarantees a clause is satisfied if and only if only one of the queries it generates can coordinate. Let us introduce some notation first. For a variable $x$, $x^1 = x$ and $x^0 = \neg x$. Moreover, define $\neg 0 = 1$ and $\neg 1 = 0$. So, an arbitrary clause over the variables $x_1, \ldots, x_k$ can be written as $C = x_{j_1}^{v_1} \vee x_{j_2}^{v_2} \vee x_{j_3}^{v_3}$. The idea is that we leave the query for the first literal, $\{R_{j_1}(v_1)\} C(1) \coloneq \emptyset$, "unconstrained". The query for the second literal is "constrained" so that it can only be "satisfied" if the first literal was not "satisfied":

$$\{R_{j_2}(v_2), R_{j_1}(\neg v_1)\} C(1) \coloneq \emptyset .$$

Similarly, the query for the third literal is "constrained" so that it can only be "satisfied" if the other two literal were not "satisfied":

$$\{R_{j_3}(v_3), R_{j_2}(\neg v_2), R_{j_1}(\neg v_1)\} C(1) \coloneq \emptyset .$$

This way, a clause is satisfied if and only if only one of its corresponding queries is "satisfied". For a clause $C$, call the set of these three entangled queries $Q(C)$. For the example query $C = x_1 \vee \neg x_2 \vee x_3$ we would get the queries:

$$\{R_1(1)\} C(1) \coloneq \emptyset$$
$$\{R_2(0), R_1(0)\} C(1) \coloneq \emptyset$$
$$\{R_3(1), R_2(1), R_1(0)\} C(1) \coloneq \emptyset$$

From a 3SAT instance $\mathcal{C} = \{C_1, \ldots, C_k\}$, we obtain an ENTANGLEDMAX instance with set of queries $Q(C_1) \cup \cdots \cup Q(C_k) \cup \{q(x_1), \ldots, q(x_m)\}$, and database $I$ that interprets $D$ as $\{0, 1\}$. For example, the 3SAT instance $C_1 = x_1 \vee$ $\neg x_2 \vee x_3$, $C_2 = x_2 \vee \neg x_3 \vee \neg x_4$ would give us queries with the coordination graph shown in Figure 9.

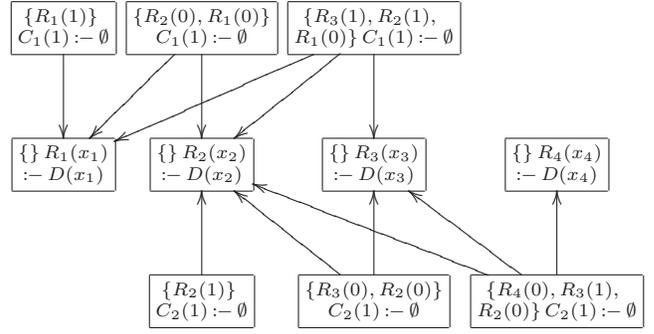

**Figure 9: Coordination Graph for the Proof of Theorem 2**

Recall that for each clause at most one of the queries associated with it can be part of a coordinating set. Hence, the size of the maximal coordinating set is at most $k + m$. Moreover, it is not hard to show that $\mathcal{C}$ is a "yes" instance of 3SAT if and only if the size of the largest coordinating set is $k + m$. ∎

## B. EXTENDING SECTION 5

We show that a relaxation of the conditions we defined in Section 5 for sets of consistent queries brings back hardness. We give a reduction from 3SAT using a construction that involves queries coordinating on different sets of attributes (some queries coordinate on an attribute $A_0$ and the rest on attributes $A_0, A_1$).

Let $\mathcal{C} = \{C_1, \ldots, C_k\}$ be a set of clauses, with each clause containing exactly 3 literals. Let $x_1, \ldots, x_n$ be the variables that appear in $\mathcal{C}$. We introduce a query $q_\mathcal{C}$ that intuitively requires all the clauses to be true:

$$(q_\mathcal{C}) : \ \{R(y_1, C_1), \ldots, R(y_k, C_k)\} \ R(x, \mathcal{C}) \coloneq$$
$$\text{Fl}(x, \mathsf{1MAR}), \bigwedge_i \text{Fl}(y_i, \mathsf{1MAR})$$

Each clause is true if at least one of its literals is true. For each variable $x_i$ we have a "positive literal" query $q_{X_i}$ and a "negative literal" query $q_{X_i^*}$.

$(q_{C_j}): \{R(y, f)\} \ R(x, C_j) \coloneq \text{Fr}(C_j, f), \text{Fl}(x, \mathsf{1MAR}), \text{Fl}(y, d)$

$(q_{X_i}): \{R(y, S_i)\} \ R(x, X_i) \coloneq \text{Fl}(x, \mathsf{1MAR}), \text{Fl}(y, \mathsf{1MAR})$

$(q_{X_i^*}): \{R(y, S_i)\} \ R(x, X_i^*) \coloneq \text{Fl}(x, \mathsf{2MAR}), \text{Fl}(y, \mathsf{2MAR})$

A clause $C_j$ has as "friends" the literals that can satisfy it. For example, if $C_j = x_1 \vee \neg x_2 \vee x_3$, then the Friends (Fr) relation in the database will contain $(C_j, X_1), (C_j, X_2^*), (C_j, X_3)$. The crucial part of the construction is a "selection gadget" that forces only one of the literals for a given variable to be able to coordinate:

$$(S_i) : \ \{R(y, \mathcal{C})\} \ R(x, S_i) \coloneq \text{Fl}(x, d), \text{Fl}(y, d')$$

Notice the circular dependency we have created: Every coordinating set will have to include the query $q_\mathcal{C}$. Variable assignments are encoded as selections of the literal queries $q_{X_i}, q_{X_i^*}$. Let $Q$ be the collection of all these queries. It is easy to see that $\mathcal{C}$ is satisfiable iff there is a coordinating set in $Q$.